\documentclass[twocolumn,groupedaddress,assymb,amsmath]{revtex4}
\usepackage{graphicx}
\usepackage{amssymb}
\usepackage{amsmath}

\usepackage{extarrows}
\usepackage{color}
\usepackage[colorlinks,citecolor=blue, linkcolor=blue,hyperindex,CJKbookmarks,dvipdfm]{hyperref}
\usepackage{lipsum} 

\begin{document}
\title{Enhancing the sensitivity of nonlinearity sensors through homodyne detection in dissipatively coupled systems}
  \author{Dianzhen Cui$^{1}$, Jianning Li$^{1}$, Fude Li$^{1}$, Zhi-Cheng Shi$^{3,4}$, X. X.  Yi$^{1,2}$\footnote{yixx@nenu.edu.cn}}
\affiliation{$^1$Center for Quantum Sciences and School of Physics, Northeast Normal University, Changchun 130024, China\\
$^2$Center for Advanced Optoelectronic Functional Materials Research, and Key Laboratory for UV Light-Emitting Materials and Technology of Ministry of Education, Northeast Normal University, Changchun 130024, China\\
$^3$Fujian Key Laboratory of Quantum Information and Quantum Optics (Fuzhou University), Fuzhou 350108, China\\
$^4$Department of Physics, Fuzhou University, Fuzhou 350108, China}

  \date{\today}
 \begin{abstract}
In this manuscript, we propose a new sensing mechanism to enhance the sensitivity of a quantum system  to nonlinearities by homodyning the amplitude quadrature of the cavity field. The system consists of two dissipatively coupled cavity modes, one of which is subject to single- and two-photon drives. In the regime of low two-photon driving strength, the spectrum of the system acquires a real spectral singularity. We find that this singularity is very sensitive to the two-photon drive and nonlinearity of the system, and compared to the previous nonlinearity sensor, the proposed sensor achieves an unprecedented sensitivity around  the singularity point. Moreover, the scheme is  robust against  fabrication imperfections. This work would open a new avenue for  quantum sensors, which could find applications in many  fields, such as the  precise measurement and quantum metrology.
\end{abstract}
\maketitle

\section{Introduction}
Hermiticity and real eigenvalues of the Hamiltonian in closed systems are the key postulate in the quantum mechanics. In recent years, it has been discovered that the axiom of Hermiticity can be replaced by the condition of parity-time ($\mathcal{PT}$) symmetry, leading to the foundations of non-Hermitian quantum mechanics \cite{Bender1998,Bender2002}. Interestingly, the non-Hermitian Hamiltonians also exhibit entirely real eigenvalues when satisfying $[H,\mathcal{PT}]=0$, where $\mathcal{PT}$ is the joint parity-time operator. A more significant feature of such Hamiltonians is the breaking of the $\mathcal{PT}$ symmetry, in which the eigenspectrum switches from purely real to completely imaginary \cite{Bender2007,Berry2004,Heiss2012,El-Ganainy2007,Guo2009,Ruter2010,Peng2014,Hang2013,Zhang2016,
Fleury2015,Yong2015,Zhu2014,Feng2014,Hodaei2014,Wiersig2014,Liu2016,
Chen2017,Ganainy2018,Doppler2016,Harris2016,Q. Wang2020,Jing2014,Hodaei2017}. This sudden $\mathcal{PT}$ phase transition is marked by the exceptional point (EP), associated with level coalescence, in which the eigenvalues and their corresponding eigenvectors simultaneously coalesce and become degenerate. Recently, the $\mathcal{PT}$ phase transition has been experimentally observed in various $\mathcal{PT}$ symmetric systems \cite{Rubinstein2007, Schindler2011,Bittner2012,Feng2011}.

As a counterpart, the anti-parity-time ($\mathcal{APT}$) symmetry, namely the Hamiltonian of the system is anti-commutative with the joint $\mathcal{PT}$ operator (mathematically, $\{H,\mathcal{PT}\}=0$), has recently attracted great interest \cite {Jiang2019,J. M. P.Nair2021,Ge2013,Peng2016,Yang2017,Konotop2018,Zhang2018,Chuang2018,Li2018,Zhao2020,
Antonosyan2015,Choi2018,Fan2020,Li2019,Zhang2020,Arkhipov2022,Wiersig2020}.  In contrast to the $\mathcal{PT}$ symmetric system, the $\mathcal{APT}$ symmetric
system does not require gain, but it can still exhibit EP with purely imaginary eigenvalues. This characteristic is of great significance for realizing non-Hermitian dynamics in the quantum domain without Langevin noise \cite{Scheel2018}. Until now, several  relevant experiments have been realized in different physical systems, including cold atoms \cite{Chuang2018}
, optics \cite{Li2019}, magnon-cavity hybrid systems \cite{Zhao2020}, electrical circuit resonators \cite{Choi2018}, and integrated photonics \cite{Yang2017,Fan2020}.

Sensitivity enhancement based on EP has been demonstrated both theoretically and experimentally \cite{Zhang2020,Ganainy2018,Wiersig2020,Wiersig2014,Liu2016,Hodaei2017,A. Alu2019,L. Yang2019,Chen2017,Djorwe2018,Djorwe2019,Tao2021,Cui2021,Mao2020,Scheel2018,Digonnet2021} in the particle detector \cite{Wiersig2014}, mass sensor \cite{Djorwe2019}, and gyroscope \cite{Mao2020}. It has been shown that if an EP is subjected
to the strength $\epsilon$ of the \textit{linear} perturbation, the frequency splitting (the energy spacing of the two levels) scales as the square root of the perturbation strength $\epsilon$ \cite{Zhang2020,Ganainy2018,Wiersig2020,Wiersig2014,Liu2016,Hodaei2017,A. Alu2019,L. Yang2019,Chen2017,Djorwe2018,Djorwe2019,Tao2021,Cui2021,Mao2020,Scheel2018,Digonnet2021}. Recently, in the context of dissipatively coupled $\mathcal{APT}$ symmetric systems, a scheme was proposed to efficiently detect the \textit{nonlinear} perturbations \cite{J. M. P.Nair2021}.
This dissipatively coupled system has an imaginary coupling strength \cite{J. M. P.Nair2021}, resulting from the fact that the vacuum of the electromagnetic field can produce coherence in the process of spontaneous emission \cite{Agarwal1974}. Owing to this coherence, the system acquires a real spectral singularity which strongly suppresses the linewidth of a resonance spectrum, thereby drawing out a remarkable  response. Particularly, near the coherence-induced singularity (CIS), the response $\mathcal{N}$ behaves as $\left|\frac{d\mathcal{N}}{d U}\right|\varpropto|U|^{-5/3}$,  where $U$ quantifies the strength of the Kerr nonlinearity \cite{J. M. P.Nair2021}. Compared with EP-based sensors \cite{Hodaei2017,Wiersig2020,Scheel2018,Wiersig2014,Liu2016,Chen2017,Ganainy2018,Zhang2020,A. Alu2019,L. Yang2019,Djorwe2018,Djorwe2019,Mao2020,Tao2021,Cui2021,Digonnet2021}, the sensitivity of the system to inherent nonlinearities has been greatly improved, and the protocol does not require any gains \cite{J. M. P.Nair2021}.

To enhance the  sensitivity of quantum sensors, in this manuscript, we theoretically propose a novel sensing mechanism to improve the sensitivity of a quantum  system to nonlinearities. Our proposal is based on two dissipatively coupled cavity modes, one of which is subject to single- and two-photon drives.  The key point of our sensing protocol is that the spectrum of the dissipatively coupled system acquires a CIS at the low two-photon driving strength. In the vicinity of the CIS, the current sensing protocol exhibits a much larger sensitivity compared with the previous work \cite{J. M. P.Nair2021}. The proposed sensor differs from known sensors in at least five points: (i) it operates at a CIS instead of the EP, (ii) it can help to  estimate  two types of nonlinear parameters (Kerr nonlinearity coefficient and  two-photon drive amplitude), (iii) the nonlinear parameters are estimated via the amplitude quadrature of the cavity field and can reach an unprecedented sensitivity around the CIS, (iv) we only need to slightly increase the two-photon driving amplitude to overcome the deleterious effects of fabrication imperfections, and (v)
it works without the requirement of  $\mathcal{APT}$ symmetry, making such configuration much more accessible.

The remainder of this manuscript is organized as follows. In Sec.~\ref{sec1}, a physical model is introduced to describe the setup, and the dynamical equations for the system are derived. In Sec.~\ref{sec2}, we study the sensitivity of the system to nonlinearities and discuss the effect of the fabrication imperfections on the performance of the setup. In Sec. ~\ref{sec3}, we discuss the experimental feasibility of the present scheme. Finally, the conclusions are drawn in Sec.~\ref{sec4}.

\section{Model and dynamical equations}\label{sec1}
\begin{figure}[h]
	\centering
	\includegraphics[width=0.48\textwidth]{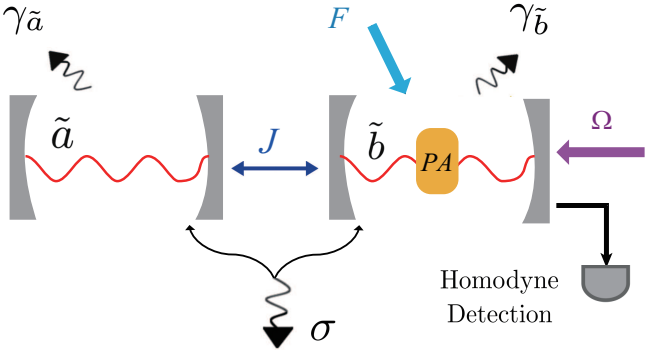}

	\caption {Illustration of our setup. The cavity modes $\tilde{a}$ and $\tilde{b}$ are
dissipatively coupled through the shared  dissipative environment. $\gamma_{\tilde{a}}$ and $\gamma_{\tilde{b}}$ denotes  dissipation rate of the system $\tilde{a}$ and $\tilde{b}$, respectively. $\sigma$ stands for the cooperative interactions between the two modes and the common reservoir. $J$ is the direct coupling between the cavity modes. The cavity mode $\tilde{b}$ is driven by a single-photon pump of amplitude $\Omega$. The cavity mode $\tilde{b}$ is also subject to a classical pump of amplitude $F$. A parametric amplifier (PA) is inserted inside the cavity $\tilde{b}$. By using a homodyne setup, the nonlinear parameter is estimated via the amplitude quadrature of the cavity field $\tilde{b}$. See the text for  details.}
	\label{Fig1}

\end{figure}

The schematic diagram is sketched in Fig. \ref{Fig1}. We consider a general situation where we have two dissipatively coupled cavity modes, one of which is subject to single- and two-photon drives. In the rotating frame with respect to  frequency $\omega_p$ of the laser, the total Hamiltonian of the system reads $(\hbar=1)$ \cite{J. M. P.Nair2021,Aspelmeyer2014},

\begin{eqnarray}
\begin{aligned}
H=H_{f}+H_k+H_{i}+H_{c}+H_{s},
\label{Hamiltonian 1}
\end{aligned}
\end{eqnarray}
with
\begin{eqnarray}
\begin{aligned}
H_f&=\Delta_{\tilde{a}} a^\dagger a+\Delta_{\tilde{b}} b^\dagger b,
\\H_k&=U(b^{\dagger 2}b^2),
\\H_{i}&=J(a b^\dag+a^\dag b),
\\H_{l}&=\Omega(b^{\dagger}+b),
\\H_{s}&=G(e^{-i\theta_p}b^{\dagger 2}+e^{i\theta_p}b^2).
\label{Hamiltonian 2}
\end{aligned}
\end{eqnarray}
Here $H_{f}$ represents the free Hamiltonian of the uncoupled cavity modes $\tilde{a}$ and $\tilde{b}$, $a^\dagger$ $(b^\dagger)$ and $a$ $(b)$ are the creation and annihilation operators of the mode $\tilde{a}$ ($\tilde{b}$), respectively. $\Delta_j=\omega_{j}-\omega_p/2$ $(j=\tilde{a},\tilde{b})$ represents the detuning of the modes $\tilde{a}$ and $\tilde{b}$ with respect to the laser field, and the frequencies of the modes $\tilde{a}$ and $\tilde{b}$ are $\omega_{\tilde{a}}$ and $\omega_{\tilde{b}}$, respectively. The Hamiltonian $H_{k}$ describes the Kerr nonlinearity of the mode $\tilde{b}$, and the strength is denoted by $U$ \cite{Bayen2019,Bayen2020,Mandal2019,Gerry1989,Buzek1989,Tanas1989}. The Hamiltonian $H_{i}$ describes the direct coupling between the modes with  coupling strength $J$. The Hamiltonian $H_{l}$ represents the mode $\tilde{b}$  driven coherently by a single-photon pump with amplitude $\Omega$ and frequency $\omega_l=\omega_p/2$. The Hamiltonian $H_{s}$ describes  the mode $\tilde{b}$ subjected to a two-photon drive of amplitude $G$, frequency $\omega_{p}$, and phase $\theta_p$. We would demonstrate later how to adjust $\theta_{p}$ to enhance the response of the system to Kerr nonlinearity. Physically, a squeezed laser can be obtained by means of the degenerate parametric down-converter \cite{Gerry2005}. A certain kind of nonlinear
medium is pumped by a field of frequency $\omega_p$ and the photons of that field
are converted into pairs of identical photons, of frequency  $\omega_p /2$ each, into the signal field. This process is known as the degenerate parametric down-conversion and it can be implemented in a system described by the Hamiltonian
\begin{equation}
\begin{aligned}
H= &\omega_{\tilde{a}} a^{\dagger} a +\omega_{\tilde{b}} b^{\dagger} b+J(a b^\dag+a^\dag b)+U b^\dag b^\dag b b\\
&+\Omega(b^{\dagger}e^{-i \omega_l t}+b e^{i \omega_l t}) +\omega_{p} c^{\dagger} c +\chi^{(2)}\left(b^{2} c^{\dagger}+b^{\dagger 2} c\right),
\label{eqr1}
\nonumber
\end{aligned}
\end{equation}
where $\omega_p$ is the frequency of the pump mode and $\chi^{ (2)}$ is a second order nonlinear susceptibility \cite{Boyd2008}. We now  assume that the pump field is  classical, such that its photons remains undepleted over the relevant time scale. Suppose that the field is in coherent state $|\Lambda e^{-i\omega_p t}\rangle$ ($\Lambda =F e^{-i\theta_p}$) and approximate the operators $c$ and $c^\dagger$ by $\Lambda e^{-i\omega_p t}$ and $\Lambda^* e^{i\omega_p t}$, respectively, the above Hamiltonian reduces to,
\begin{equation}
\begin{aligned}
H= &\omega_{\tilde{a}} a^{\dagger} a +\omega_{\tilde{b}} b^{\dagger} b+J(a b^\dag+a^\dag b)+U b^\dag b^\dag b b\\
&+\Omega(b^{\dagger}e^{-i \omega_l t}+b e^{i \omega_l t}) +\left(\tilde{G} ^*b^{2} e^{i\omega_p t} + \tilde{G} b^{\dagger 2}  e^{-i\omega_p t} \right),
\nonumber
\end{aligned}
\end{equation}
where $\tilde{G}=\chi^{(2)} \Lambda=G e^{-i\theta_p}$.
In the rotating frame defined by $U=\exp \left[\left(-i \frac{\omega_p}{2} a^{\dagger} a-i \frac{\omega_p}{2} b^{\dagger} b\right) t\right]$, the above Hamiltonian becomes
\begin{equation}
\begin{aligned}
H= &\Delta_{\tilde{a}} a^{\dagger} a+ \Delta_{\tilde{b}} b^{\dagger} b+J(a b^\dag+a^\dag b)+U b^\dag b^\dag b b\\
&+ \Omega(b^{\dagger}e^{-i \Delta_l t}+b e^{i \Delta_l t})+ \left(\tilde{G} ^*b^{2} +\tilde{G} b^{\dagger 2} \right),
\nonumber
\end{aligned}
\end{equation}
where $\Delta_{\tilde{a}/\tilde{b}}=\omega_{\tilde{a}/\tilde{b}}-\omega_p/2$ and $\Delta_{l}=\omega_{l}-\omega_{p}/2$.   Recently, this two-photon drive has been realized by coupling two superconducting resonators through a Josephson junction \cite{Leghtas2015}. On the other side, the dissipative environment can be roughly divided into two categories--one is that the modes are coupled independently to their local reservoirs, and the other is that a common reservoir interacts with both, as shown in Fig. \ref{Fig1}.

A complete description of the two-mode system interacting with the dissipative environment is  the master equation in the Lindblad form \cite{Agarwal1974,Metelmann2015},
\begin{eqnarray}
\begin{aligned}
\frac{d\rho}{dt}=&-i[H,\rho]+\gamma_{\tilde{a}} \mathcal{L}\left[a\right]\rho+\gamma_{\tilde{b}} \mathcal{L}\left[b\right]\rho+\sigma \mathcal{L}\left[c\right]\rho,
\label{master equation}
\end{aligned}
\end{eqnarray}
where the second and third terms represent the intrinsic damping of the modes $\tilde{a}$ and $\tilde{b}$, respectively. The fourth term describes the cooperative interactions between the two modes and the common reservoir. The standard dissipative superoperator $\mathcal{L}[o]$ is defined by $\mathcal{L}[o]\rho=2o \rho o^\dagger-o^\dagger o \rho-\rho o^\dagger o$, and the jump operator $c$ is a linear superposition of the annihilation operators $a$ and $b$,  $c \rightarrow \nu a+u e^{i\theta}b$. If the phase difference $\theta$ of light propagation from one mode to another is a multiple of $2 \pi$, the jump operator has the general form  $c \rightarrow \nu a+u b$ \cite{Metelmann2015}, where the coefficients $\nu$ and $u$ represent the couplings of the two modes to the common reservoir, respectively. If the two modes are symmetrically coupled to the common reservoir, the operator $c$ is expressed as $c=(1/\sqrt{2})(a+b)$. The external damping rates induced by the common reservoir for the two modes are $\sigma\cdot\nu^2=\kappa_{\tilde{a}}$ and $\sigma\cdot u^2=\kappa_{\tilde{b}}$, respectively. The cooperative dissipations between the two modes is $\sigma\cdot\nu u=\sqrt{\kappa_{\tilde{a}}\kappa_{\tilde{b}}}$, where the $\sqrt{\kappa_{\tilde{a}}\kappa_{\tilde{b}}}$ represents the effect of quantum interference resulting from the cross coupling between the two modes. Without loss of generality, we assume that the parameters $\gamma_{j}$ and $\kappa_{j}$ $(j=\tilde{a},\tilde{b})$ are the same for the whole system, i.e., $\gamma_{\tilde{a}}=\gamma_{\tilde{b}}=\gamma_0$, and $\kappa_{\tilde{a}}=\kappa_{\tilde{b}}=\Gamma$.

\section{Sensitivity At The coherence-induced singularity}\label{sec2}
\subsection{Effective Hamiltonian and the sensitivity of the system to nonlinearities}
Starting from the Lindblad master equation in Eq. (\ref{master equation}), we can obtain the mean value equations for the modes $\tilde{a}$ and $\tilde{b}$ via the relation $\langle \dot{\zeta} \rangle=Tr(\dot{\rho}\zeta)$ \cite{Agarwal1974}

\begin{eqnarray}
\begin{aligned}
\dot{\alpha}=&-i[\Delta_{\tilde{a}}-i(\gamma_0+\Gamma)]\alpha-i(J-i\Gamma)\beta,
\\ \dot{\beta}=&-i(J-i\Gamma)\alpha-i[\Delta_{\tilde{b}}-i(\gamma_0+\Gamma)]\beta\\
&-2iU |\beta|^2 \beta-2iG \beta^*-i\Omega,
\\ \dot{\alpha^*}=&\big(\dot{\alpha}\big)^*,
\\ \dot{\beta^*}=&\big(\dot{\beta}\big)^*,
\label{QLEs1}
\end{aligned}
\end{eqnarray}
where $\dot{\alpha}=\left\langle \dot{a} \right\rangle$, $\dot{\beta}=\langle \dot{b} \rangle$ and we set $\theta_p=0$. In the derivation of Eq. (\ref{QLEs1}), we have adopted the mean field approximation, i.e., $\left\langle b^\dagger b b\right\rangle\approx
\left\langle b^\dagger\right\rangle \left\langle b \right\rangle \left\langle b \right\rangle$. In the next sections, we work in the parameter interval, in which the mean field approximation is a good approximation. It is obvious to observe from the above expressions that the effective dissipative coupling strength between the two modes is $i\Gamma$, which originates from the bath-mediated collective damping.

To study the  performance of the proposed sensor, we need to solve the eigenvalues of the effective optical system and find the CIS feature. The equivalent
Schr\"{o}dinger-like equation in this configuration obeys $i\frac{d\phi}{dt}=H_{eff}\phi$,
where $\phi=(\alpha,\beta,\alpha^*,\beta^*)^T$ is the state vector, and the form of the associated effective Hamiltonian $H_{eff}$ is,
\begin{widetext}
\begin{equation}
H_{eff}=\left(
 \begin{array}{cccc}
 \Delta_{\tilde{a}}-i(\gamma_0+\Gamma)&J-i\Gamma&0&0\\
 J-i\Gamma&\Delta_{\tilde{b}}-i(\gamma_0+\Gamma)+2\tilde{U}&0&2G\\
 0&0&-\Delta_{\tilde{a}}-i(\gamma_0+\Gamma)&-J-i\Gamma\\
 0&-2 G &-J-i\Gamma&-\Delta_{\tilde{b}}-i(\gamma_0+\Gamma)-2\tilde{U}
 \end{array}
\right),
\label{effective Hamiltonian1}
\end{equation}
\end{widetext}
where $\tilde{U}=U|\beta|^2$. Notably, the effective Hamiltonian (\ref{effective Hamiltonian1}) does not have the $\mathcal{APT}$ symmetry. Therefore, our system is easier to obtain than the previous schemes \cite{J. M. P.Nair2021,Zhang2020}. The effective Hamiltonian $H_{eff}$ has four eigenvalues forming two pairs and one pair is due to the appearance of $\alpha^*$ and $\beta^*$ in the dynamics.

In the limit of the weak two-photon driving amplitude $G$, we can bring the effective  Hamiltonian into a block diagonal form, and  we will  study the block corresponding to $\alpha$ and $\beta$ in the following,
\begin{eqnarray}
\begin{aligned}
\tilde{H}_{eff}= \left(
 \begin{array}{cc}
 \Delta_{\tilde{a}}-i(\gamma_0+\Gamma)&J-i\Gamma\\
 J-i\Gamma&\tilde{\Delta}_{\tilde{b}}-i(\gamma_0+\Gamma)
 \end{array}
\right).
\label{effective Hamiltonian}
\end{aligned}
\end{eqnarray}
where $\tilde{\Delta}_{\tilde{b}}=\Delta_{\tilde{b}}+2\tilde{U}$.
Without loss of generality, we choose  the   parameter as follows, $\Delta_{\tilde{a}}=-\Delta_{\tilde{b}}=\delta/2$, $J=0$, and $\tilde{U}=10^{-3}$ $\Gamma$, similar to
the parameters chosen in Ref. \cite{J. M. P.Nair2021}.
The eigenvalues of Eq. (\ref{effective Hamiltonian}) are given by
\begin{eqnarray}
\begin{aligned}
\tilde{\lambda}_{\pm}&=\tilde{U}- i (\gamma_0+\Gamma)\pm\frac{1}{2}\sqrt{4 \tilde{U}^{2}-4 \Gamma^{2}-4 \tilde{U} \delta+\delta^{2}}
\\&\approx -i(\gamma_0+\Gamma)\pm\sqrt{\frac{\delta^2}{4}-\Gamma^2}.
\label{analytical}
\end{aligned}
\end{eqnarray}
With the intrinsic damping $\gamma_0$ of the mode approaching zero, one of the eigenvalues characterizing its dynamics tends  to the real axis at $\delta=0$. The dissipative coupling strength $i\Gamma$ can be viewed as an effective gain that
offsets exactly the external dissipation of the coupled resonance.
\begin{figure}[h]
	\centering

	\includegraphics[width=0.52\textwidth]{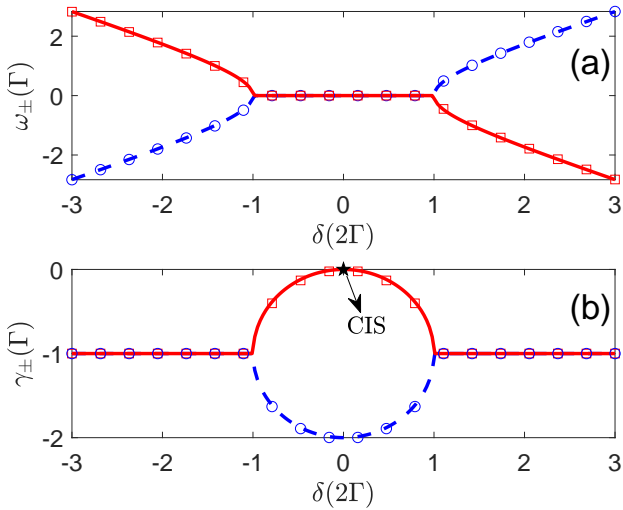}

	\caption {(a) The real and (b) the imaginary parts of the eigenvalues $vs$ the detuning $\delta$ with $\gamma_0=10^{-5}\Gamma$. The circles and squares are the numerical results with $G=0.01\Gamma$ and $\tilde{U}=10^{-3}\Gamma$, while the solid and dashed lines represent the analytical approximation given in Eq. (\ref{analytical}). The CIS is marked by the star. We only study the eigenvalue corresponding to $\alpha$ and $\beta$.}
	\label{Fig2}

\end{figure}

The solid and dashed lines in Fig. \ref{Fig2} (a) and (b) show the real ($\omega$) and imaginary parts ($\gamma$) of the eigenvalues [see Eq. (\ref{analytical})] as a function of the detuning $\delta$. For comparison, we numerically solve the eigenvalues of Eq. (\ref{effective Hamiltonian1}) at $G=0.01\Gamma$ and $\tilde{U}=10^{-3}\Gamma$ [see the circles and squares in Fig. \ref{Fig2} (a) and (b)]. We see that the numerical and analytical results are highly agreement at a weak two-photon driving amplitude $G$. This validates the approximations we made in the calculations. Figure \ref{Fig2} (b) shows the spectrum of the dissipatively coupled system acquires a CIS in the limit $\delta=0$ and $\gamma_0\rightarrow0$. The extreme condition $\gamma_0=0$ holds when none of the cavity modes suffers spontaneous losses from the surroundings while interacting with the mediating bath. The CIS has prodigious sensing potential, allowing efficient detection of nonlinearities in the configuration \cite{J. M. P.Nair2021}. The physical origin of this peculiar behavior comes from an effective coupling induced between two modes in the presence of a shared reservoir.

The CIS was exploited to measure the response (mean excitation number for the system in steady-state) of the system to the parameter change of the Kerr nonlinearity with only a single-photon drive in Ref.  \cite{J. M. P.Nair2021}. Here, we elaborate a novel detection strategy through homodyne detection. Specifically, we perform a homodyne measurement on the cavity field $\tilde{b}$ to detect the weak nonlinearities with higher sensitivity. The key measurement quantity, in this case, is the amplitude and phase quadratures of the cavity field. Solving the steady-state solutions of Eq. (\ref{QLEs1}), we obtain
\begin{eqnarray}
\begin{aligned}
-i[\delta/2-i\gamma]\alpha&-\Gamma\beta=0,
\\ -\Gamma\alpha-i[-\delta/2-i\gamma]\beta&-2iU |\beta|^2 \beta-2i G \beta^*=i\Omega,
\\ i[\delta/2+i\gamma]\alpha&^*-\Gamma\beta^*=0,
\\ -\Gamma\alpha^*+i[-\delta/2+i\gamma]\beta&^*+2iU |\beta|^2 \beta^*+2i G \beta=-i\Omega.
\label{QLEs2}
\end{aligned}
\end{eqnarray}
Eliminating $\alpha$ and $\alpha^*$, we get
\begin{small}
\begin{eqnarray}
\begin{aligned}
\frac{\Gamma^2 \beta}{i\delta/2+\gamma}+(i\delta/2-\gamma)\beta-2iU |\beta|^2 \beta-2iG \beta^*&=i\Omega,\\
\frac{\Gamma^2 \beta^*}{-i\delta/2+\gamma}+(-i\delta/2-\gamma)\beta^*+2iU |\beta|^2 \beta^*+2iG \beta&=-i\Omega,
\label{ steady state relations}
\end{aligned}
\end{eqnarray}
\end{small}
where $\gamma=\gamma_0+\Gamma$. Defining the amplitude  $X_{\tilde{b}}=\frac{b+b^\dag}{\sqrt{2}}$, and the phase $  Y_{\tilde{b}}=\frac{-i\left(b-b^\dag\right)}{\sqrt{2}}$, the expressions of the amplitude and phase quadratures of the cavity mode $\tilde{b}$ is given by
\begin{widetext}
\begin{eqnarray}
\begin{aligned}
\langle  X_{\tilde{b}}\rangle=&\frac{\sqrt{2}\Omega}{-\frac{\Theta^2\gamma^2}
{\left(\gamma^2+(\delta/2)^2\right)^2\left(\frac{(\delta/2)\Gamma^2}{\gamma^2+(\delta/2)^2}-\delta/2 +2U|\beta|^2-2G\right)}-\left(\frac{(\delta/2)\Gamma^2}{\gamma^2+(\delta/2)^2}-\delta/2 +2U |\beta|^2+2G\right)},\\
\langle  Y_{\tilde{b}}\rangle=&\frac{\sqrt{2}\Omega}{\frac{\Theta\gamma}{\gamma^2+(\delta/2)^2}+\frac{\gamma^2+(\delta/2)^2}{\Theta\gamma}
\left[\left(\frac{(\delta/2)\Gamma^2}{\gamma^2+(\delta/2)^2}-\delta/2 +2U |\beta|^2+2G\right)\left(\frac{(\delta/2)\Gamma^2}{\gamma^2+(\delta/2)^2}-\delta/2 +2U |\beta|^2-2G\right)\right]},
\label{homodyning}
\end{aligned}
\end{eqnarray}
\end{widetext}
where $\Theta=\Gamma^2-\gamma^2-(\delta/2)^2$. Especially, $\Theta$ becomes extremely small around the CIS, which will cause $\langle  Y_{\tilde{b}}\rangle$ to converge to 0. However, the expression of the amplitude quadrature of the cavity field $\tilde{b}$ around the CIS can be further simplified as

\begin{eqnarray}
\begin{aligned}
\left|\langle  X_{\tilde{b}}\rangle\right|\approx \frac{\Omega}{\sqrt{2}(U |\beta|^2+G)}.
\label{amplitude01}
\end{aligned}
\end{eqnarray}
\begin{figure}[h]
	\centering

	\includegraphics[width=0.5\textwidth]{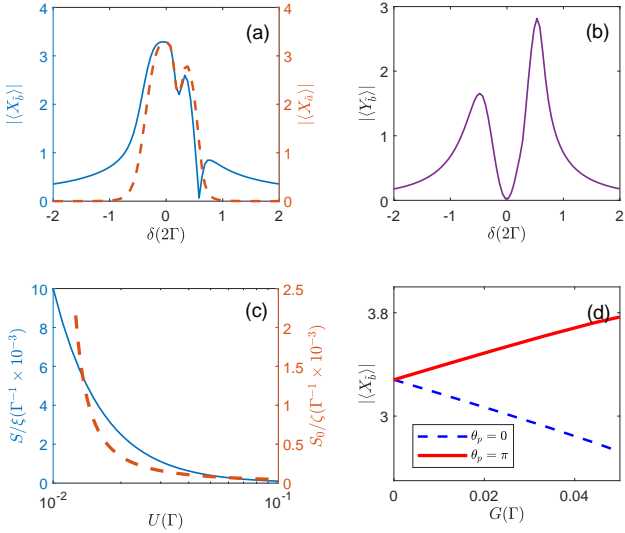}

	\caption {(a) Average of the amplitude quadrature $|\langle  X_{\tilde{b}}\rangle|$ and $|\langle  X_{\tilde{a}}\rangle|$ for the system in steady-state $vs$ the detuning $\delta$ at $U=0.02\Gamma$ and $G=0$. (b) $|\langle  Y_{\tilde{b}}\rangle|$ $vs$ $\delta$ at $U=0.02\Gamma$ and $G=0$. (c) Contrast between the normalized sensitivity $S/\xi$ and $S_0/\zeta$ when the $U$ is below $0.1\Gamma$ at $G=0$. Notice that the coordinate scales for $S$ and $S_0$ are different. The orange axis indicates the product of the $S_0/\zeta$ and $10^{-3}$, while the blue axis indicates the product of the $S/\xi$ and $10^{-3}$. (d) $|\langle  X_{\tilde{b}}\rangle|$ $vs$ $G$ around the CIS at $U=0.02\Gamma$.
The other system parameters are the same as in Fig. \ref{Fig2}.}
	\label{Fig3}

\end{figure}
The introduction of the two-photon drive reduce the sensitivity of the sensor, nevertheless, we can eliminate this influence by setting $G\ll \tilde{U}$. In this situation, we can approximately obtain
\begin{eqnarray}
\begin{aligned}
\left|\langle  X_{\tilde{b}}\rangle\right|\approx \frac{\Omega}{\sqrt{2}U |\beta|^2}.
\label{amplitude1}
\end{aligned}
\end{eqnarray}
Clearly, Eq. (\ref{amplitude1}) shows an excellent nonlinear dependence of the amplitude $\left|\langle  X_{\tilde{b}}\rangle\right|$ on the Kerr nonlinear coefficient $U$ around the CIS. To validate the superiority of utilizing CIS, we numerically plot the amplitude and phase averages of the cavity field $\tilde{b}$ as a function of the detuning $\delta$ in Fig. \ref{Fig3} (a) and (b). In the absence of two-photon drive, the amplitude average $|\langle  X_{\tilde{b}}\rangle|$ displays a sharp peak to $U$ around the CIS. A similar result was obtained by homodyning the amplitude of the cavity field $\tilde{a}$. This suggests  that CIS is a useful tool for sensing the Kerr nonlinearity. In contrast to the amplitude average, the phase average get a dip near the CIS, as predicted by the second line of Eq. (\ref{homodyning}). To this extent, we can choose to measure the amplitude quadrature to estimate the Kerr nonlinear coefficient $U$ . The corresponding sensitivity quantitatively characterizes the  performance of the sensor operating at CIS. The sensitivity can be defined as
\begin{eqnarray}
\begin{aligned}
S=\left|\frac{d\langle  X_{\tilde{b}}\rangle}{d U}\right|= \xi U^{-2},
\label{sensitivity1}
\end{aligned}
\end{eqnarray}
where $\xi=\Omega/(\sqrt{2}|\beta|^2)$. In order to reveal the advantages of our sensing mechanism, a comparison with previous sensing protocol is necessary. For the previous nonlinearity sensor, the sensitivity was expressed as $S_0=\zeta U^{-5/3}$ \cite{J. M. P.Nair2021}. Figure \ref{Fig3} (c) shows the normalized sensitivity $S$ and $S_0$ versus $U$. The sensitivity of the proposed sensor has been considerably enhanced in comparison with the previous nonlinearity sensor. In addition, we note that the tuning of the phase $\theta_ p$ of the two-photon drive plays an important role in enhancing the response of the system to Kerr nonlinearity. When we set $\theta_ p=\pi$, Eq. (\ref{amplitude01})
becomes
\begin{eqnarray}
\begin{aligned}
\left|\langle  X_{\tilde{b}}\rangle\right|\approx \left|\frac{\Omega}{\sqrt{2}(U |\beta|^2-G)}\right|,
\end{aligned}
\end{eqnarray}
where the sign of the two-photon driving amplitude $G$ is flipped.
It means that the response of the system to Kerr nonlinearity can be enhanced to some extent by increasing the two-photon driving amplitude $G$. This analysis is in agrement with our numerical  results given in  Fig. \ref{Fig3} (d).

\begin{figure}[h]
	\centering

	\includegraphics[width=0.52\textwidth]{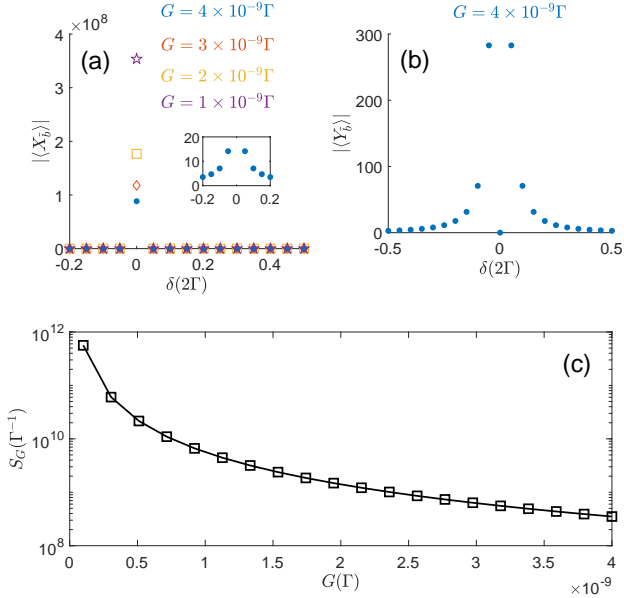}

	\caption {Average of (a) the amplitude quadrature $|\langle  X_{\tilde{b}}\rangle|$  and (b) the phase quadrature $|\langle  Y_{\tilde{b}}\rangle|$ for the system in steady-state $vs$ the detuning $\delta$ at four different strengths of the two-photon drive. A significant response of the system to two-photon drive can be found around the CIS, otherwise the response of the system to two-photon drive was weak (see the inset). (c) The sensitivity $S_G$ $vs$ $G$ around the CIS. The other system
parameters are the same as in Fig. \ref{Fig2}.}
	\label{Fig4}

\end{figure}

Another important finding of our work is that the system is also highly sensitive to two-photon drive. We consider the case where the Kerr nonlinearity is weak with respect to the two-photon drive, at which point the Kerr nonlinearity can be safely ignored. In Fig. \ref{Fig4} (a) and (b), we plot the amplitude and phase averages of the cavity field $\tilde{b}$ as a function of the detuning $\delta$ at four different strengths of the two-photon drive. Similarly, the amplitude average shows a striking response to $U$ around the CIS. A weaker nonlinearity begets a higher response, as manifested in Fig. \ref{Fig4} (a). And the phase average tends to 0 around the CIS [see Fig. \ref{Fig4} (b)].
In this case, the sensitivity of the system to the two-photon drive is obtained as follows,
\begin{eqnarray}
\begin{aligned}
S_G=\left|\frac{d\langle  X_{\tilde{b}}\rangle}{d G}\right|\propto G^{-2}.
\label{sensitivity}
\end{aligned}
\end{eqnarray}
Clearly, near the CIS, the amplitude average becomes drastically sensitive to variations in $G$ [see Fig. \ref{Fig4} (c)], proving the efficiency of the CIS-based sensor in detecting the two-photon drive. Thus, our work provides a new way of estimating the two-photon driving amplitude for the CIS-based sensor.

\begin{figure}[h]
	\centering

	\includegraphics[width=0.53\textwidth]{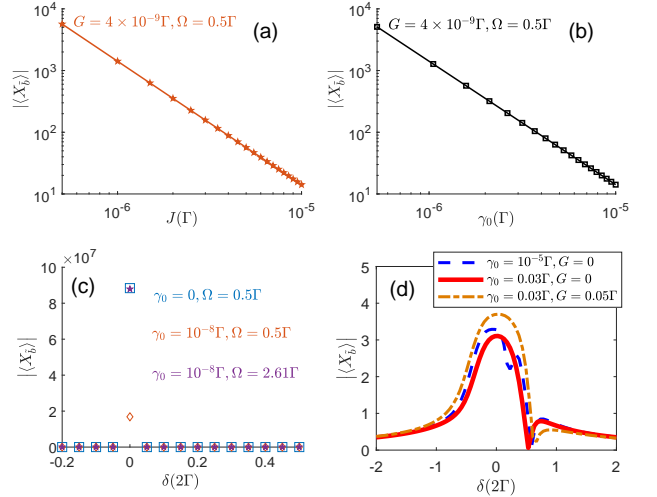}

	\caption {(a) The response of the system to two-photon drive as a function of the direct coupling $J$ around the CIS at $\gamma_0=0$. (b) The response of the system to two-photon drive $vs$ the intrinsic damping $\gamma_0$ around the CIS. (c) The response of the system to two-photon drive $vs$ the detuning $\delta$ at $G=4\times10^{-9}\Gamma$. (d) The response of the system to Kerr nonlinearity as a function of the detuning $\delta$ at $\theta_p=\pi$, $\Omega=0.5\Gamma$, and $U=0.02\Gamma$.}
	\label{Fig5}

\end{figure}

\subsection{The effect of fabrication imperfections on the sensitivity}
The present scheme works for zero cavity-cavity couplings and ignorable intrinsic dampings.  In realistic scenarios, however, fabrication imperfections are unavoidable. In this section, we investigate the effects of fabrication  imperfections on the performance of the scheme. Firstly, for a system consisting of two cavities with the non-zero coupling $J$ ($J \ll \Gamma$). The eigenvalues of Eq. (\ref{analytical}) become $\tilde{\lambda}_{\pm}=-i\Gamma\pm\sqrt{(J-i\Gamma)^2+\delta^2/4}$ at $\gamma_0=0$. We get a near-CIS around $\delta = 0$. Note that the amplitude and phase quadratures of Eq. (\ref{homodyning}) is now modified by replacing $\Gamma^2$ with $-(J-i\Gamma)^2$.
We plot the modified response (amplitude average) of the system to two-photon
drive as a function of the direct coupling $J$ around the CIS [see Fig. \ref{Fig5} (a)]. The introduction of the direct coupling $J$ between the two modes results in a decrease in the amplitude average, disrupting the performance of the CIS-based sensor. Secondly, the nonzero $\gamma_0$ (with a finite linewidth) also results in a decrease in the response [see Fig. \ref{Fig5} (b)], and this decrease can be cancelled by appropriately increasing the single-photon driving amplitude $\Omega$ [see Fig. \ref{Fig5} (c)]. A single-photon driving amplitude close to 2.61 $\Gamma$ returns an amplitude average corresponding to zero intrinsic damping. Similar conclusions can be obtained for the sensing of kerr nonlinearity. In contrast to the sensing of the two-photon drive, we only need to slightly increase the two-photon driving amplitude to overcome this deleterious effects of fabrication imperfections, as revealed in Fig. \ref{Fig5} (d). In this sense, the present work provides a new way for CIS-based sensor that is robust against  defects or fabrication imperfections.

\section{Discussion of experimental feasibility}\label{sec3}
\begin{figure}[h]
	\centering

	\includegraphics[width=0.4\textwidth]{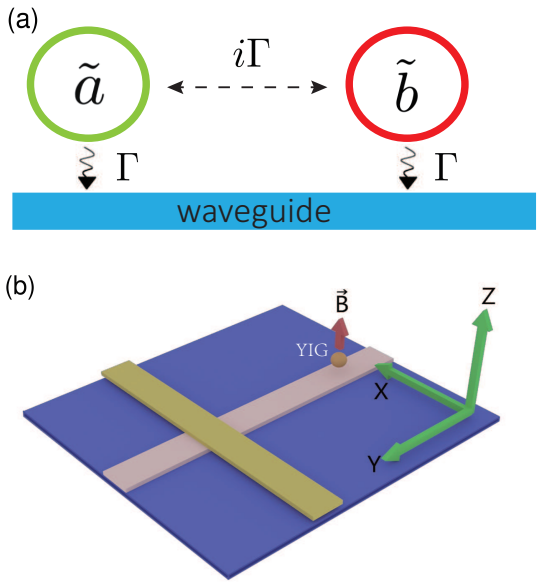}

	\caption {(a) Schematic of the dual-resonator system that demonstrates CIS. The two micro-ring resonators are dissipatively coupled through a 1D waveguide. The effective dissipative coupling strength between the two modes is $i\Gamma$. (b) Schematic of the cavity-magnonic setup. The device consists of a yttrium iron garnet (YIG) sphere and a cross-line microwave circuit. A external  magnetic field $\vec{B}$ aligned along the $Z$ axis produces the Kittel mode in YIG.}
	\label{Fig6}

\end{figure}

Owing to recent progress in nanofabrication, our sensing protocol can be  realized  in experiments \cite{Gong2021,X. Sun2022}. Here, we consider a silicon integrated photonic apparatus comprising two micro-ring resonators, both of them are coupled to a one-dimensional (1D) waveguide, as depicted in Fig. \ref{Fig6} (a). The two micro-ring resonators have a radius of 3.1 $\mu$m and the waveguide width of 0.4 $\mu$m. The gaps between the micro-ring resonators and the waveguide are 0.1 $\mu$m. In the setup, the silicon has a negligible intrinsic damping in the communication band ($\sim1550$ nm). The resonant frequency can be tuned precisely by the electro-optic effects. Owing to the large distance between the two resonators, the direct coupling between them can be ignored. Thus, such a configuration constitutes a good benchmark to test our protocol.

To make our protocol work, a driving laser of the frequency $\omega_l$ is applied to the mode $\tilde{b}$, the field amplitudes are given by $\Omega=\sqrt{\kappa_jP_l/\hbar\omega_l}$ ($j=\tilde{a},\tilde{b}$), where $P_l$ is the input laser power. A two-photon drive can be realized via optical parametric down-conversion.
Within the reach of current experiment \cite{Gong2021}, the system parameters in this study can be chosen as $\Gamma=1334$ GHz and $Q\sim10^4$ ($Q$-factor). These together with the attainable  two-photon driving amplitude   $G\in[133.4,5336] $Hz,  the response of the system to two-photon drive falls in the region of $(3.536-0.088)$ $\times 10^9$.

We would like to mention that  our  protocol is not limited to  this particular architecture. For example, it can be realized in the cavity-magnon configurations repored in \cite{Agarwal084001,You224410,M. Harder2018,YPWang2019,Hu2020,Yang09154,Rao2020}, where we  consider a setup consists of a microwave cavity and a YIG sphere, both interfacing with a 1D waveguide [see Fig. \ref{Fig6} (b)]. The microwave cavity is subject to a two-photon drive of amplitude $G$ and a microwave field with the Rabi frequency $\Omega$. Due to the absence of spatial overlap between the optical cavity and magnon modes, the direct coupling
between them can be safely ignored. The interaction with the waveguide inducing a dissipative magnon-photon coupling with $\Gamma=2\pi\times 10$ MHz. With an achievable two-photon driving amplitude range $G\in[0.0063,0.2513] $ Hz, our sensing protocol theoretically predicts that the response of the system to two-photon drive falls in the range of $(3.536-0.088)$ $\times 10^9$. It can also be seen that the response of the system to two-photon drive is greatly increased for weak two-photon driving amplitude. This validates the efficiency of the CIS-based sensor in detecting weak nonlinearities.

\section{Conclusion}\label{sec4}

In conclusion, we have proposed a new mechanism to enhance the sensitivity of the system to nonlinearities by homodyning the amplitude quadrature of the cavity field.  The  system consists of two dissipatively coupled cavity modes, one of which is  subject to single- and two-photon drives. For low two-photon driving strength, the spectrum of the dissipatively coupled system acquires a CIS, which exhibits high sensitivity to weak nonlinearities. The physical origin of this peculiar behavior lies in the effective coupling induced between two modes in the presence of a common reservoir. Compared to the previous sensing protocol, the sensor achieves an unprecedented sensitivity around the CIS. We
illustrate the sensing capabilities in two scenarios, one with a silicon integrated photonic apparatus and the other with a cavity-magnonic setup. Our scheme is robust against the fluctuations and open new avenue for weak nonlinearities. It is worth  noting that our scheme does not require the $\mathcal{APT}$ symmetric prerequisite, and can be extended to a plethora of systems, including, laser-cooled atomic ensembles \cite{Jiang2019}, superconducting transmon qubits \cite{Koch2007}, optomechanical systems \cite{Bernier2018,Yu2021,Q. Zhang2021,Stannigel2010}. Despite here we focus on estimating a nonlinear parameter, our sensing protocol, in principle, can also be applied to
estimate the linear parameter.

  \section{acknowledgments}\label{sec5}
This work is supported by National Natural Science Foundation of China (NSFC) under Grants No. $12175033$ and No. $12147206$  and National Key R$\&$D Program of China under Grant No. 2021YFE0193500.

\end{document}